\documentclass[%
 reprint,
 amsmath,amssymb,
 aps,prd,
 floatfix,
]{revtex4-2}

\usepackage[T1]{fontenc}
\usepackage{lmodern}
\usepackage{mathtools}
\usepackage{graphicx}
\usepackage{xcolor}
\usepackage{microtype}
\usepackage{xurl}
\usepackage{hyperref}
\usepackage{orcidlink}
\hypersetup{
 colorlinks=true,
 linkcolor=blue!55!black,
 citecolor=blue!55!black,
 urlcolor=blue!55!black,
 pdfauthor={Abdullah Guvendi, Omar Mustafa and Claudio Furtado},
 pdftitle={Exact scalarization thresholds and onset laws for a nonminimally coupled scalar field on the Ellis wormhole},
 bookmarksopen=true,
 bookmarksnumbered=true,
 pdfstartview=FitH
}
\allowdisplaybreaks[2]

\newtheorem{proposition}{Proposition}

\begin{document}

\title{Exact scalarization thresholds and onset laws for a nonminimally coupled scalar field on the Ellis wormhole}

\author{Abdullah Guvendi\orcidlink{0000-0003-0564-9899}}
\email{abdullah.guvendi@erzurum.edu.tr}
\affiliation{Department of Basic Sciences, Erzurum Technical University, 25050 Erzurum, T\"urkiye}

\author{Omar Mustafa\orcidlink{0000-0001-6664-3859}}
\email{omar.mustafa@emu.edu.tr}
\affiliation{Department of Physics, Eastern Mediterranean University, 99628 G. Magusa, North Cyprus, Mersin 10, T\"urkiye}

\author{Claudio Furtado\orcidlink{0000-0002-3455-4285}}
\thanks{Corresponding author}
\email{furtado@fisica.ufpb.br}
\affiliation{Departamento de F\'isica, Universidade Federal da Para\'iba, Caixa Postal 5008, Jo\~ao Pessoa, Para\'iba, 58051-970, Brazil}

\date{\today}

\begin{abstract}
We study a real scalar field nonminimally coupled to curvature on a fixed Ellis wormhole background. The scalar perturbation decouples at linear order, and the substitution $x=a\tan z$ reduces its static radial equation to a trigonometric P\"oschl--Teller problem. This yields the exact threshold spectrum $\zeta_{n\ell}=(n+\ell+1)^2/2$, closed-form Gegenbauer profiles, and an exact count of negative radial eigenvalues in each angular sector. The $s$-wave thresholds are zero-energy resonances at which the growth rate rises linearly, whereas the $\ell\geq1$ threshold states are square integrable and the growth rate rises as a square root. At the first threshold we also obtain the next-order term, $\sigma a=(\pi/4)g-(3\pi/16)g^2+O(g^3)$ with $g=2\zeta-1$. For a probe field with quartic self-interaction, we show that no nontrivial spherical static solution exists for $\zeta\leq1/2$ and derive local branch expansions near the $s$-wave thresholds. Numerical evolutions are consistent with saturation on the nodeless branch. These nonlinear results refer to the probe-field model; they do not establish the existence or stability of fully backreacted scalarized wormholes.
\end{abstract}

\maketitle

\section{Introduction}\label{sec:intro}

In spontaneous scalarization, a solution with vanishing scalar field becomes unstable once the coupling between the scalar and curvature (or matter) crosses a critical value. Nonlinear terms may then halt the growth and support solutions with a nonzero scalar field, although whether such branches exist, and whether they are stable, has to be established model by model. The mechanism was discovered for neutron stars in scalar--tensor theories \cite{DamourEspositoFarese1993} and later extended to black holes coupled to the Gauss--Bonnet invariant \cite{DonevaYazadjiev2018,Silva2018,Antoniou2018}; Ref.~\cite{DonevaRMP2024} reviews the subject. The linear onset is usually located by looking for a static solution of the scalar equation in which curvature acts as an effective mass. Such a zero mode marks a candidate bifurcation point, and the existence of a nonlinear branch must then be shown separately.

Wormholes are a natural, if less explored, setting for this mechanism. Under the usual assumptions, traversable wormholes require matter that violates the null energy condition \cite{MorrisThorne1988,MorrisThorneYurtsever1988,Visser1989,VisserBook,HochbergVisser1997,HochbergVisser1998,FordRoman1996}. This alone does not force the Ricci scalar to be nonzero, but for the Ellis geometry it is, whereas it vanishes for the vacuum black holes of general relativity. Scalarized wormholes have already been built in Einstein--scalar--Gauss--Bonnet gravity \cite{AntoniouWH2020,BrihayeRenaux2020,Ibadov2021}. We focus on the Ellis wormhole \cite{Ellis1973,Bronnikov1973}: smooth, free of horizons, ultrastatic, and asymptotically flat on both sides of the throat. Two of its properties single it out for Ricci-type scalarization. Its Gauss--Bonnet invariant vanishes identically (Sec.~\ref{sec:geometry}), so the channel that scalarizes black holes is closed. Its Ricci scalar, by contrast, is negative everywhere and is fixed on shell by the trace of the supporting matter. The coupling $\zeta R\Phi^2$ is therefore the wormhole analog of the coupling to the trace of stellar matter that scalarizes neutron stars \cite{DamourEspositoFarese1993}.

The linear problem is that of a massless scalar with nonminimal coupling on the fixed Ellis background. Taylor showed that, for a class of wormholes that includes Ellis, the $s$-wave becomes unstable when $\zeta>1/2$ \cite{Taylor2014}. Quasinormal modes, scalar clouds, and the self-force on Ellis--Bronnikov geometries have also been studied \cite{BlazquezChewKunz2018,Azad2023,BlazquezSalcedo2022,Taylor2013,BezerraKhusnutdinov2009,KhusnutdinovBakhmatov2007}. Of particular relevance here, Ref.~\cite{BlazquezSalcedo2022} found that a minimally coupled scalar admits no smooth, localized linear clouds on this background. With a nonminimal coupling the situation is different. Smooth static clouds exist, but only at a discrete set of couplings that we compute exactly. Each of these thresholds marks the onset of a new unstable mode in the scalar sector; whether every threshold also gives rise to a branch of fully backreacted solutions is a separate nonlinear question.

Three features of the problem allow us to go well beyond a generic test-field calculation. First, $\Phi$ enters the gravitational and matter equations only quadratically, so its linear perturbations around the scalar-free Ellis solution decouple from those of the metric and of the phantom field. The test-field spectrum is then the exact $\Phi$ sector of the full linear problem (Sec.~\ref{sec:backreaction}). Second, the substitution $x=a\tan z$ turns the radial equation into a trigonometric P\"oschl--Teller problem. This yields the whole tower of thresholds, and, once the roles of coupling and frequency are exchanged, it also yields a compact Sturm--Liouville problem for the growing modes. Third, in that compact problem the threshold eigenvalues are isolated. The birth of each unstable mode can therefore be treated by ordinary perturbation theory, even though in the original variables the threshold sits at the edge of the continuous spectrum.

Our results are:
\begin{enumerate}
\item the full tower of thresholds $\zeta_{n\ell}=(n+\ell+1)^2/2$, with Gegenbauer profiles in closed form, together with their classification as zero-energy resonances ($\ell=0$) or normalizable zero modes ($\ell\geq1$);
\item an exact compact formulation of the growing-mode problem, which proves by a monotonicity argument that sector $\ell$ has $N_\ell(\zeta)=\max\{0,\lceil\sqrt{2\zeta}\,\rceil-\ell-1\}$ unstable modes and also provides an efficient numerical method;
\item the onset law at every threshold: linear growth with a closed-form coefficient at the $s$-wave resonances, square-root growth with a beta-function coefficient for $\ell\geq1$, and the exact second-order term at the first threshold;
\item a nonlinear probe-field model with quartic self-interaction, for which we prove a no-hair result in the spherical sector for $\zeta\leq1/2$, derive the leading branch amplitudes and energy near the lowest thresholds, and present time evolutions consistent with saturation on the nodeless branch.
\end{enumerate}

Our results concern a scalar field on a prescribed geometry and do not require the background itself to be dynamically stable. The Ellis spatial geometry also appears outside gravitation. Its equatorial section is the catenoid \cite{Dandoloff2010}, which has been studied in curved graphene \cite{deSouza2022} and in nematic optical media \cite{Azevedo2021}, in the spirit of analog gravity \cite{BarceloLRR2011,LeonhardtPhilbin2006}. In these systems the geometry is fixed by construction, so a probe-field treatment is exactly what is needed. The Ellis Ricci scalar is, moreover, twice the Gaussian curvature $K=-a^2/(x^2+a^2)^2$ of the catenoid written in terms of arclength $x$, the quantity that sets the geometry-induced potential on that surface \cite{Dandoloff2010}.

Throughout the paper we keep three kinds of statements apart: exact results for the linear scalar operator, numerical evidence for the nonlinear probe model, and conjectures about fully backreacted solutions. The distinction matters because the supporting phantom field has an instability of its own, which the scalar-sector calculation does not remove.

The paper is organized as follows. Section~\ref{sec:geometry} describes the geometry and its curvature invariants, and Sec.~\ref{sec:radial} derives the radial operator. Section~\ref{sec:threshold} solves the static problem. Section~\ref{sec:compact} sets up the compact formulation and counts the unstable modes, and Sec.~\ref{sec:onset} derives the onset laws, which Sec.~\ref{sec:numerics} checks numerically. Section~\ref{sec:nonlinear} treats nonlinear scalarization in the probe limit, and Sec.~\ref{sec:backreaction} relates the results to the fully coupled theory. We conclude in Sec.~\ref{sec:conclusion}. The Appendices give the beta-function sums and the numerical details. We use units with $c=\hbar=G=1$.

\section{Geometry and curvature invariants}\label{sec:geometry}

In proper radial coordinates the Ellis wormhole reads
\begin{equation}
ds^2=-dt^2+dx^2+r^2(x)\,d\Omega_2^2,\qquad r^2(x)=x^2+a^2,
\label{eq:metric}
\end{equation}
where $x\in\mathbb R$, the throat of radius $a>0$ sits at $x=0$, and the signature is $(-,+,+,+)$. Our curvature conventions are $R^\rho{}_{\sigma\mu\nu}=\partial_\mu\Gamma^\rho_{\nu\sigma}-\partial_\nu\Gamma^\rho_{\mu\sigma}+\Gamma^\rho_{\mu\lambda}\Gamma^\lambda_{\nu\sigma}-\Gamma^\rho_{\nu\lambda}\Gamma^\lambda_{\mu\sigma}$ and $R_{\sigma\nu}=R^\rho{}_{\sigma\rho\nu}$. Since the metric is ultrastatic, every curvature component with a time index vanishes, and the spacetime curvature reduces to that of the spatial warped product. The Ricci scalar is
\begin{equation}
R=-\frac{4r''}{r}+\frac{2(1-r'^2)}{r^2}=-\frac{2a^2}{(x^2+a^2)^2}.
\label{eq:ricci}
\end{equation}
It is smooth and even in $x$, negative everywhere, equal to $-2/a^2$ at the throat, and falls off as $|x|^{-4}$.

\emph{Gauss--Bonnet invariant.} By the same argument, $\mathcal G=R^2-4R_{\mu\nu}R^{\mu\nu}+R_{\mu\nu\rho\sigma}R^{\mu\nu\rho\sigma}$ equals the same combination built from the curvature of the three-dimensional spatial slices. The Weyl tensor vanishes in three dimensions, so $R_{ijkl}R^{ijkl}=4R_{ij}R^{ij}-R^2$, and therefore
\begin{equation}
\mathcal G\equiv0.
\label{eq:GB}
\end{equation}
This holds for any ultrastatic metric, whose spatial slices are static and have vanishing extrinsic curvature. A coupling $f(\Phi)\mathcal G$ therefore provides no local source on this background, and among quadratic-curvature couplings the Ricci coupling is the relevant one.

\emph{Relation to the supporting matter.} In general relativity the trace of the Einstein equations gives $R=-8\pi T$. For the Ellis solution supported by a phantom scalar, $T=a^2/[4\pi(x^2+a^2)^2]$, so that $\zeta R\Phi^2=-8\pi\zeta\,T\Phi^2$. The scalar thus couples to the trace of the exotic matter, in direct analogy with the matter coupling that scalarizes neutron stars \cite{DamourEspositoFarese1993}.

\section{Field equation and radial operator}\label{sec:radial}

We take the action
\begin{align}
S_\Phi=&-\frac12\int d^4x\sqrt{-g}\,\Big(g^{\mu\nu}\nabla_\mu\Phi\nabla_\nu\Phi+\zeta R\Phi^2\Big)\nonumber\\
&-\int d^4x\sqrt{-g}\,\frac{\lambda}{4}\Phi^4,
\label{eq:action}
\end{align}
with $\lambda\geq0$ \cite{BirrellDavies1982,Faraoni1996}. The quartic term matters only in Sec.~\ref{sec:nonlinear}. Sections~\ref{sec:radial}--\ref{sec:numerics} deal with the linear equation
\begin{equation}
(\Box-\zeta R)\Phi=0.
\label{eq:kg}
\end{equation}
Writing $\Phi=r^{-1}u_\ell(t,x)Y_{\ell m}(\theta,\phi)$ and using $r^{-2}\partial_x[r^2\partial_x(u/r)]=r^{-1}(u''-r''u/r)$, we obtain the reduced wave equation $\partial_t^2u_\ell-\partial_x^2u_\ell+V_\ell u_\ell=0$ with the exact potential
\begin{equation}
V_\ell(x)=\frac{\ell(\ell+1)}{x^2+a^2}+(1-2\zeta)\frac{a^2}{(x^2+a^2)^2}.
\label{eq:potential}
\end{equation}
For $u_\ell=e^{-i\omega t}\psi_\ell(x)$ this becomes $H_\ell(\zeta)\psi_\ell=\omega^2\psi_\ell$, with $H_\ell(\zeta)=-d^2/dx^2+V_\ell$. The potential is smooth and bounded and vanishes at both ends. Hence $H_\ell$ is self-adjoint on $H^2(\mathbb R)\subset L^2(\mathbb R)$, and Weyl's theorem gives $\sigma_{\rm ess}(H_\ell)=[0,\infty)$ \cite{ReedSimon1975,ReedSimonIV}. A negative eigenvalue $\omega^2=-\sigma^2$ corresponds to a normalizable mode that grows as $e^{\sigma t}$.

When $\zeta\leq1/2$ the potential is nonnegative and there are no negative eigenvalues. When $\zeta>1/2$ the $s$-wave potential is $-gW(x)$, with $g=2\zeta-1$ and $W=a^2/(x^2+a^2)^2\in L^1$. In one dimension any nonpositive, integrable potential that is not identically zero has a bound state \cite{Simon1976}, so the instability sets in exactly at $\zeta_c=1/2$, in agreement with Ref.~\cite{Taylor2014}. The rest of the paper describes what happens above this value.

\section{Static solutions and the threshold tower}\label{sec:threshold}

For a static field $\Phi=\phi_\ell(x)Y_{\ell m}$, Eq.~\eqref{eq:kg} reads
\begin{equation}
\frac{d}{dx}\!\left[(x^2+a^2)\frac{d\phi_\ell}{dx}\right]-\ell(\ell+1)\phi_\ell+\frac{2\zeta a^2}{x^2+a^2}\phi_\ell=0.
\label{eq:staticx}
\end{equation}
We introduce the compact coordinate
\begin{equation}
x=a\tan z,\qquad z\in(-\pi/2,\pi/2),
\label{eq:z}
\end{equation}
which sends the two asymptotic ends to $z=\pm\pi/2$. Equation~\eqref{eq:staticx} becomes exactly the trigonometric P\"oschl--Teller problem \cite{PoschlTeller1933}
\begin{equation}
P_\ell\phi\equiv\left[-\frac{d^2}{dz^2}+\ell(\ell+1)\sec^2z\right]\phi=2\zeta\,\phi.
\label{eq:pt}
\end{equation}
Setting $p=\ell+1$, $y=\sin z$, and $\phi=(1-y^2)^{p/2}F(y)$ leads to the Gegenbauer equation $(1-y^2)F''-(2p+1)yF'+(2\zeta-p^2)F=0$, whose solutions are regular at both endpoints only if $2\zeta-p^2=n(n+2p)$ \cite{Szego1975,NIST2010}. The thresholds and their profiles are therefore
\begin{align}
\zeta_{n\ell}&=\frac{(n+\ell+1)^2}{2},\qquad n,\ell=0,1,2,\ldots,\label{eq:thresholds}\\
\phi_{n\ell}(z)&=(\cos z)^{\ell+1}\,C_n^{\ell+1}(\sin z).\label{eq:gegenbauer}
\end{align}

\emph{Endpoint behavior.} The corresponding Schr\"odinger function is $\psi_{n\ell}=r\phi_{n\ell}=a(\cos z)^\ell C_n^{\ell+1}(\sin z)$. Because $C_n^{\ell+1}(\pm1)\neq0$, it behaves as $\psi_{n\ell}\sim|x|^{-\ell}$ at large $|x|$. For $\ell=0$ it approaches a nonzero constant at both ends; it is bounded but not normalizable, a \emph{zero-energy resonance}. For $\ell\geq1$ it is square integrable, a genuine \emph{zero-energy eigenfunction} at the edge of $\sigma_{\rm ess}$. The field itself falls off as $\Phi\sim|x|^{-\ell-1}$. For $\ell=0$ this is a $1/|x|$ tail; its coefficient can be read as a scalar charge, but the resonance itself is not a normalizable mode. These smooth static clouds exist only at the couplings~\eqref{eq:thresholds}; for minimal coupling there are none \cite{BlazquezSalcedo2022}.

\section{Growing modes in the compact coordinate and the mode count}\label{sec:compact}

The coordinate $z$ is just as useful away from zero frequency. Substituting $\Phi=e^{\sigma t}\phi(z)Y_{\ell m}$ with $\sigma>0$ into Eq.~\eqref{eq:kg} and repeating the steps above, we find the exact equation
\begin{align}
P_\ell(\sigma)\phi&\equiv\left[-\frac{d^2}{dz^2}+\ell(\ell+1)\sec^2z+\sigma^2a^2\sec^4z\right]\phi\nonumber\\
&=2\zeta\,\phi .
\label{eq:compactkappa}
\end{align}
Here the coupling $2\zeta$ is the eigenvalue, and the growth rate appears as the strength of a confining potential. The norms are related by
\begin{equation}
\int_{\mathbb R}|\psi|^2dx=a^3\int_{-\pi/2}^{\pi/2}\sec^4z\,|\phi|^2dz,
\label{eq:norm}
\end{equation}
so a normalizable growing mode of $H_\ell(\zeta)$ is the same thing as an eigenfunction of $P_\ell(\sigma)$ with eigenvalue $2\zeta$. For $\sigma>0$ the term $\sigma^2a^2\sec^4z$ diverges as $(\pi/2\mp z)^{-4}$ at the endpoints. Both endpoints are then of limit-point type, the decaying solutions behave as $\exp[-\sigma a/(\pi/2\mp z)]$ (that is, $\psi\sim e^{-\sigma|x|}$), and the spectrum of $P_\ell(\sigma)$ is discrete and simple,
\begin{equation}
\Lambda_{0\ell}(\sigma)<\Lambda_{1\ell}(\sigma)<\Lambda_{2\ell}(\sigma)<\cdots\to\infty .
\end{equation}

\begin{proposition}[Spectral flow]\label{prop:flow}
For every $\ell$ and $n$, $\Lambda_{n\ell}(\sigma)$ is continuous and strictly increasing for $\sigma>0$. It tends to $(n+\ell+1)^2$ as $\sigma\downarrow0$ and diverges as $\sigma\to\infty$. Consequently $H_\ell(\zeta)$ has a negative eigenvalue $-\sigma^2$ on the $n$th branch if and only if $2\zeta=\Lambda_{n\ell}(\sigma)$, and this equation has a unique solution $\sigma_{n\ell}(\zeta)>0$ exactly when $\zeta>\zeta_{n\ell}$.
\end{proposition}

\emph{Proof sketch.} For $\sigma>0$ the eigenvalues are isolated and simple, and the Hellmann--Feynman theorem gives
\begin{equation}
\frac{d\Lambda_{n\ell}}{d(\sigma^2)}=a^2\frac{\int\sec^4z\,|\phi_{n\ell}^{(\sigma)}|^2dz}{\int|\phi_{n\ell}^{(\sigma)}|^2dz}>0.
\label{eq:HF}
\end{equation}
As $\sigma\downarrow0$, the quadratic forms of $P_\ell(\sigma)$ decrease monotonically to the closure of the form of $P_\ell$ on $C_c^\infty$. For $\ell=0$ this is the Dirichlet form. For $\ell\geq1$ there is only one self-adjoint realization, because $\ell(\ell+1)\geq3/4$ makes both endpoints limit point. Monotone convergence of forms \cite{Kato1995} and the min--max principle then give $\Lambda_{n\ell}(\sigma)\to\Lambda_{n\ell}(0)=(n+\ell+1)^2$, the eigenvalues found in Sec.~\ref{sec:threshold}. Divergence as $\sigma\to\infty$ follows from $\sec^4z\geq1$.

\begin{figure}[t]
\centering
\includegraphics[width=0.7\columnwidth]{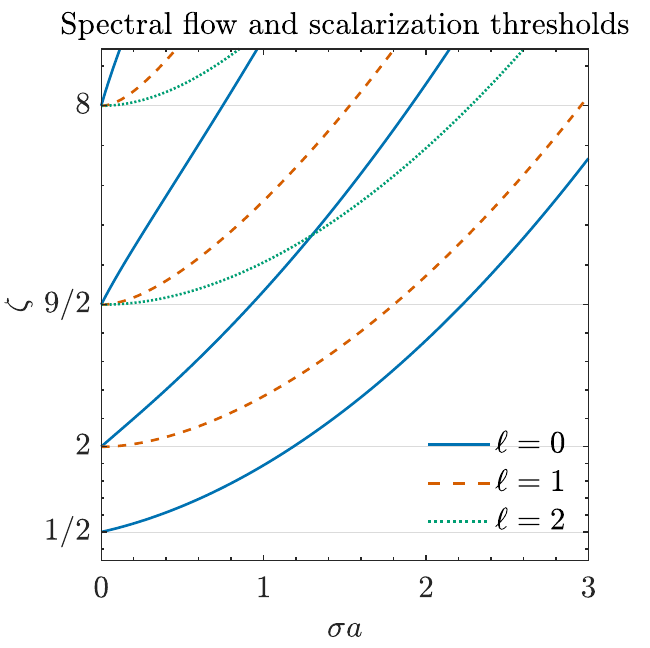}
\caption{Spectral flow of the unstable modes. Each curve gives the coupling $\zeta=\Lambda_{n\ell}(\sigma)/2$ at which a mode with growth rate $\sigma$ exists, for $\ell=0$ (solid), $1$ (dashed), and $2$ (dotted), computed from the compact problem~\eqref{eq:compactkappa}. The branches start at the thresholds $\zeta_{n\ell}=(n+\ell+1)^2/2$ (horizontal lines). They leave with finite slope for $\ell=0$ (linear onset) and with zero slope for $\ell\geq1$ (square-root onset).}
\label{fig:flow}
\end{figure}

Proposition~\ref{prop:flow} gives the number of unstable radial modes in sector $\ell$,
\begin{align}
N_\ell(\zeta)&=\#\{n\geq0:(n+\ell+1)^2<2\zeta\}\nonumber\\
&=\max\big\{0,\lceil\sqrt{2\zeta}\,\rceil-\ell-1\big\},
\label{eq:count}
\end{align}
each with multiplicity $2\ell+1$. The inequality is strict because a threshold state does not itself grow. The same count follows from the Sturm oscillation theorem on the full line \cite{ReedSimonIV,Teschl2014,GesztesySimonTeschl1996}. The subordinate zero-energy solution $u=r\phi$ has the same interior zeros as the solution of Eq.~\eqref{eq:pt} that satisfies the left endpoint condition, and the number of those zeros equals the number of eigenvalues of $P_\ell$ below $2\zeta$. Figure~\ref{fig:flow} shows the branches $\zeta=\Lambda_{n\ell}(\sigma)/2$. At a given $\zeta$, $N_\ell$ is simply the number of sector-$\ell$ branches that a horizontal line crosses.

\section{Onset laws}\label{sec:onset}

In the compact formulation the two kinds of threshold behave very differently. To expand $\Lambda_{n\ell}(\sigma)$ around $\sigma=0$ one needs the matrix element of the perturbation $\sigma^2a^2\sec^4z$ in the threshold state, $\int\sec^4z\,\phi_{n\ell}^2dz=\int\cos^{2\ell-2}z\,[C_n^{\ell+1}(\sin z)]^2dz$. This integral is finite for $\ell\geq1$ and diverges for $\ell=0$, which is the distinction between eigenvalues and resonances of Sec.~\ref{sec:threshold} seen from a different angle.

\subsection{Normalizable thresholds ($\ell\geq1$): square-root onset}

For $\ell\geq1$ the threshold state belongs to the form domain of $\sec^4z$, and $(n+\ell+1)^2$ is an isolated eigenvalue of $P_\ell(0)$. First-order asymptotic perturbation theory for monotone families of forms \cite{Kato1995,ReedSimonIV} then gives $\Lambda_{n\ell}(\sigma)=(n+\ell+1)^2+\sigma^2a^2\,I^{(-)}_{n\ell}/I^{(+)}_{n\ell}+o(\sigma^2)$, with
\begin{equation}
I_{n\ell}^{(\pm)}=\int_{-\pi/2}^{\pi/2}\cos^{2\ell\pm2}z\,[C_n^{\ell+1}(\sin z)]^2dz .
\end{equation}
Solving $2\zeta=\Lambda_{n\ell}(\sigma)$ for $\sigma$,
\begin{equation}
\sigma_{n\ell}^2=s_{n\ell}\,\frac{\zeta-\zeta_{n\ell}}{a^2}+o(\zeta-\zeta_{n\ell}),\qquad s_{n\ell}=\frac{2I^{(+)}_{n\ell}}{I^{(-)}_{n\ell}} .
\label{eq:sqrtonset}
\end{equation}
Equivalently, $-s_{n\ell}/a^2=\langle\psi_{n\ell},(\partial_\zeta H_\ell)\psi_{n\ell}\rangle/\langle\psi_{n\ell},\psi_{n\ell}\rangle$. The curvature-weighted matrix element on the threshold profile is thus exactly the slope of the eigenvalue branch, $E_{n\ell}=-\sigma_{n\ell}^2\simeq-(s_{n\ell}/a^2)(\zeta-\zeta_{n\ell})$. This is not obvious in the original variables, where the threshold lies at the edge of $\sigma_{\rm ess}$ \cite{KlausSimon1980,JensenKato1979}; it is the compact formulation that makes the eigenvalue isolated. The coefficients reduce to finite sums of beta functions (Appendix~\ref{app:beta}). For the lowest branches,
\begin{align}
s_{0\ell}&=\frac{2(\ell^2-\tfrac14)}{\ell(\ell+1)},\label{eq:s0l}\\
s_{1\ell}&:\ \tfrac14,\ \tfrac58,\ \tfrac78,\ \tfrac{21}{20},\ \tfrac{33}{28}\quad(\ell=1,\ldots,5),\nonumber
\end{align}
so that $s_{0\ell}=3/4,\,5/4,\,35/24,\,63/40$ for $\ell=1,\ldots,4$. Numerically, the correction to Eq.~\eqref{eq:sqrtonset} is of order $(\zeta-\zeta_{n\ell})^{3/2}$ for $\ell=1$ and of order $(\zeta-\zeta_{n\ell})^{2}$ for $\ell\geq2$ (Appendix~\ref{app:numerics}). We attribute the slower convergence at $\ell=1$ to the marginal decay $\psi\sim|x|^{-1}$ of its threshold states.

\subsection{Resonant thresholds ($\ell=0$): linear onset}

For $\ell=0$ the expansion is not analytic, and we obtain it by matching. Let $\psi_c$ be the bounded threshold solution at $\zeta_c=\zeta_{n0}$, with limits $\psi_\pm=\lim_{x\to\pm\infty}\psi_c$, and let $\psi$ be the growing mode at $\zeta>\zeta_c$, with $H_0\psi=-\sigma^2\psi$. Since $V_0(\zeta)-V_0(\zeta_c)=-2(\zeta-\zeta_c)W$, integrating the Wronskian identity over $\mathbb R$ gives
\begin{equation}
\sigma^2\int\psi_c\psi\,dx=2(\zeta-\zeta_c)\int W\psi_c\psi\,dx .
\end{equation}
At leading order $\psi\simeq\psi_c$ near the throat and $\psi\simeq\psi_\pm e^{-\sigma|x|}$ far from it. Then $\int\psi_c\psi\,dx\simeq(\psi_+^2+\psi_-^2)/\sigma$, and
\begin{equation}
\sigma\simeq\frac{2(\zeta-\zeta_c)\int W\psi_c^2\,dx}{\psi_+^2+\psi_-^2}.
\label{eq:resonanceonset}
\end{equation}
This is the one-dimensional resonance mechanism analyzed by Klaus and Simon \cite{KlausSimon1980,Klaus1977}. For the Ellis thresholds, $\phi_{n0}\propto\sin[(n+1)(z+\pi/2)]$, which gives $\int W\psi_c^2dx=a\int\phi_{n0}^2dz=\pi a/2$ and $\psi_\pm^2=a^2(n+1)^2$. Hence
\begin{equation}
\sigma_{n0}\,a=\frac{\pi}{2(n+1)^2}\,(\zeta-\zeta_{n0})+O\big((\zeta-\zeta_{n0})^2\big)
\label{eq:linearonset}
\end{equation}
at every $s$-wave threshold. At the first threshold we can go one order further using Simon's weak-coupling expansion \cite{Simon1976}, $\sqrt{-E}=\frac{g}{2}\int W-\frac{g^2}{4}\iint W(x)|x-y|W(y)\,dx\,dy+O(g^3)$. With $\int_{-\infty}^{\infty}|x-y|W(y)\,dy=1+(x/a)\arctan(x/a)$, the double integral evaluates to $3\pi/(4a)$, and
\begin{equation}
\sigma_{00}\,a=\frac{\pi}{4}g-\frac{3\pi}{16}g^2+O(g^3),\qquad g=2\zeta-1.
\label{eq:secondorder}
\end{equation}

\emph{Physical consequence.} Unstable modes appear far more slowly at the $s$-wave resonances than at the normalizable thresholds. At a distance $\zeta-\zeta_{n\ell}=10^{-2}$ above threshold, for example, $\sigma a\approx1.6\times10^{-2}$ for $(\ell,n)=(0,0)$ and $\sigma a\approx8.7\times10^{-2}$ for $(1,0)$, and the ratio between the two grows as $(\zeta-\zeta_{n\ell})^{-1/2}$ closer to threshold (Fig.~\ref{fig:onset}).

\section{Numerical checks}\label{sec:numerics}

\begin{figure}[t]
\centering
\includegraphics[width=\columnwidth]{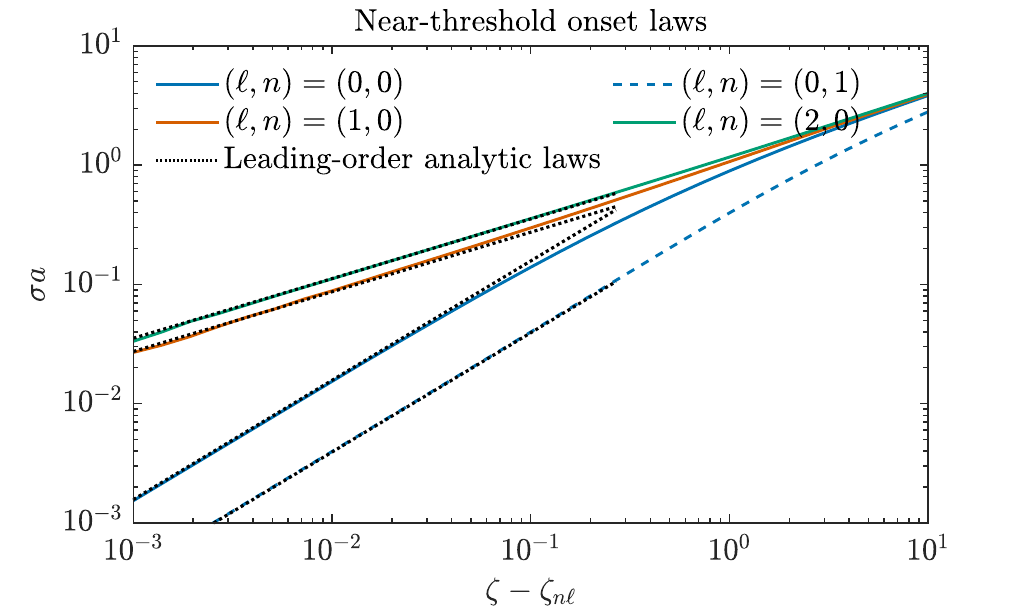}
\caption{Dimensionless growth rate $\sigma a$ as a function of the distance from threshold, $\Delta\zeta=\zeta-\zeta_{n\ell}$, for representative branches. Dotted curves show the leading asymptotic laws, Eqs.~\eqref{eq:linearonset} and \eqref{eq:sqrtonset}; they are not expected to hold far from threshold.}
\label{fig:onset}
\end{figure}

The numerical results below test the analytic formulas; they do not replace the proofs. We solve the compact problem~\eqref{eq:compactkappa} by Chebyshev collocation on the interior Gauss--Lobatto nodes \cite{Trefethen2000}. At $\sigma=0$ the method reproduces $(n+\ell+1)^2$ to machine precision, and the eigenvalues are converged to at least ten digits for $N\gtrsim120$ nodes. As an independent test we diagonalize $H_\ell$ on the full line with finite differences ($h=0.01a$, $|x|\leq150a$). At $2\zeta=\Lambda_{00}(0.5/a)=1.9140618$ this gives $E=-0.250001/a^2$, and at $2\zeta=\Lambda_{02}(1/a)=10.490369$ it gives $E=-1.00002/a^2$. Table~\ref{tab:onset} compares the onset coefficients with Eqs.~\eqref{eq:sqrtonset} and \eqref{eq:linearonset}, and Fig.~\ref{fig:onset} shows the complete growth-rate curves. At the first threshold, the second-order coefficient of $\sigma a$ in $\zeta-\tfrac12$ extracted from the numerics is $-2.3\pm0.1$, in agreement with $-3\pi/4\simeq-2.356$ from Eq.~\eqref{eq:secondorder}.

\begin{table}[t]
\caption{Leading onset coefficients. For $\ell=0$, $c=\lim_{\Delta\zeta\downarrow0}\sigma a/\Delta\zeta=\pi/[2(n+1)^2]$; for $\ell\geq1$, $c=\lim_{\Delta\zeta\downarrow0}\sigma^2a^2/\Delta\zeta=s_{n\ell}$. Numerical values are obtained from the compact problem, by Richardson extrapolation for $\ell=0$ and at $\Delta\zeta=10^{-6}$ for $\ell\geq1$.}
\label{tab:onset}
\begin{ruledtabular}
\begin{tabular}{cccccc}
$(\ell,n)$ & $\zeta_{n\ell}$ & onset & $c$ exact & (value) & $c$ numerical\\
\hline
$(0,0)$ & $1/2$ & linear & $\pi/2$ & 1.570796 & 1.570779\\
$(0,1)$ & $2$ & linear & $\pi/8$ & 0.392699 & 0.392679\\
$(0,2)$ & $9/2$ & linear & $\pi/18$ & 0.174533 & 0.174565\\
$(1,0)$ & $2$ & sqrt & $3/4$ & 0.750000 & 0.750415\\
$(1,1)$ & $9/2$ & sqrt & $1/4$ & 0.250000 & 0.250159\\
$(2,0)$ & $9/2$ & sqrt & $5/4$ & 1.250000 & 1.249999\\
$(2,1)$ & $8$ & sqrt & $5/8$ & 0.625000 & 0.625001\\
$(3,0)$ & $8$ & sqrt & $35/24$ & 1.458333 & 1.458334\\
\end{tabular}
\end{ruledtabular}
\end{table}

\begin{figure}[t]
\centering
\includegraphics[width=0.9\columnwidth]{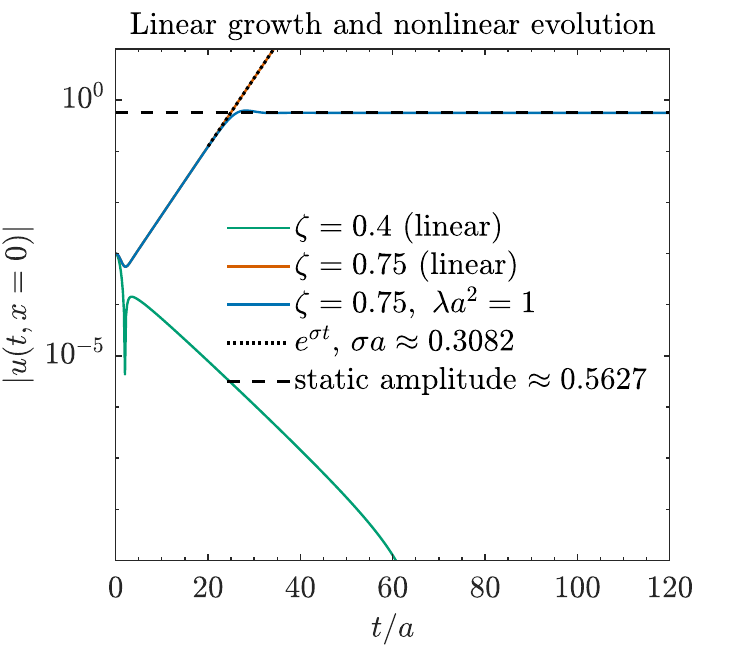}
\caption{Time evolution of the $s$-wave field at the throat, starting from a small Gaussian pulse. At $\zeta=0.4$ the perturbation decays. At $\zeta=0.75$ with $\lambda=0$ it grows at the rate predicted by the spectral problem (dotted line). For $\lambda a^2=1$ the evolution approaches the static probe-field amplitude of Sec.~\ref{sec:nonlinear} (dashed line, $\sqrt\lambda\,a\,\Phi=0.5627$). This is evidence of saturation for the chosen initial data, not a general stability proof.}
\label{fig:evol}
\end{figure}

To link these spectral results to the dynamics, we evolve the $s$-wave equation $\partial_t^2u=\partial_x^2u-V_0u-\lambda u^3/r^2$ from a small Gaussian pulse. We use fourth-order Runge--Kutta on a grid with $h=0.05a$ and $|x|\leq600a$, large enough that reflections from the boundary never reach the throat during the run. For $\zeta=0.4$ the field decays. For $\zeta=0.75$ and $\lambda=0$ it grows at the rate $0.30824/a$, while the compact problem gives $\sigma_{00}(0.75)=0.30822/a$ (Fig.~\ref{fig:evol}). We return to the nonlinear run below.

\section{Nonlinear scalarization in the probe limit}\label{sec:nonlinear}

For $\lambda>0$, a static, spherically symmetric configuration $\Phi=\phi(z)$ has energy $E[\phi]=2\pi a\,\mathcal F[\phi]$, with
\begin{equation}
\mathcal F[\phi]=\int_{-\pi/2}^{\pi/2}\!\Big[\phi_z^2-2\zeta\phi^2+\frac{\lambda a^2}{2}\sec^4z\,\phi^4\Big]dz .
\label{eq:energy}
\end{equation}
Its stationary points solve the nonlinear eigenvalue problem
\begin{equation}
-\phi''+\lambda a^2\sec^4z\,\phi^3=2\zeta\phi,\qquad \phi(\pm\pi/2)=0 .
\label{eq:nonlinear}
\end{equation}
The weight $\sec^4z$ that confines the growing modes in Eq.~\eqref{eq:compactkappa} now multiplies the self-interaction.

\begin{proposition}[Probe-limit scalarization]\label{prop:nonlinear}
In the spherical sector of the probe-field model: (i) for $\zeta\leq1/2$, every admissible finite-energy weak solution of Eq.~\eqref{eq:nonlinear} is trivial; (ii) for $\zeta>1/2$, the energy~\eqref{eq:energy} takes negative values and is bounded below; (iii) if local branches bifurcate from the $s$-wave thresholds in a suitable weighted function space, their leading amplitudes, obtained by projection onto the threshold profiles $n=0$ and $n=1$, are
\begin{equation}
A_0^2=\frac{2\zeta-1}{2\lambda a^2},\qquad A_1^2=\frac{2\zeta-4}{12\lambda a^2},
\label{eq:pitchfork}
\end{equation}
and the nodeless branch has energy
\begin{equation}
E_0\simeq-\frac{\pi^2(2\zeta-1)^2}{4\lambda a}.
\end{equation}
\end{proposition}

\emph{Proof and scope.} (i) Multiplying Eq.~\eqref{eq:nonlinear} by $\phi$ and integrating gives $\int\phi'^2+\lambda a^2\int\sec^4z\,\phi^4=2\zeta\int\phi^2$ for every admissible finite-energy weak solution for which the integration is justified. The Dirichlet Poincar\'e inequality, $\int\phi'^2\geq\int\phi^2$, then forces $\phi=0$ when $2\zeta\leq1$ and $\lambda>0$. (ii) For $\zeta>1/2$ the trial profile $\epsilon\cos z$ gives $\mathcal F=\tfrac{\pi}{2}\epsilon^2(1-2\zeta)+O(\epsilon^4)<0$ for small nonzero $\epsilon$, while the quartic term bounds $\mathcal F$ from below. A complete existence proof by the direct method would require specifying the completion of the weighted energy space and establishing compactness and lower semicontinuity there. (iii) Projecting onto $s_n(z)=\sin[(n+1)(z+\pi/2)]$, with $\int\sec^4z\,s_0^4\,dz=\pi$ and $\int\sec^4z\,s_1^4\,dz=6\pi$, gives the leading coefficients~\eqref{eq:pitchfork}, and inserting $A_0$ into Eq.~\eqref{eq:energy} gives $E_0$. Because $\sec^4z$ is singular at the endpoints, the projection alone does not verify the hypotheses of the Crandall--Rabinowitz theorem \cite{CrandallRabinowitz1971} in a specified Banach space. We therefore regard the amplitudes and the energy as local asymptotic coefficients, and the numerical continuation as evidence for the corresponding branches, not as a complete proof of bifurcation.

\begin{figure}[t]
\centering
\includegraphics[width=\columnwidth]{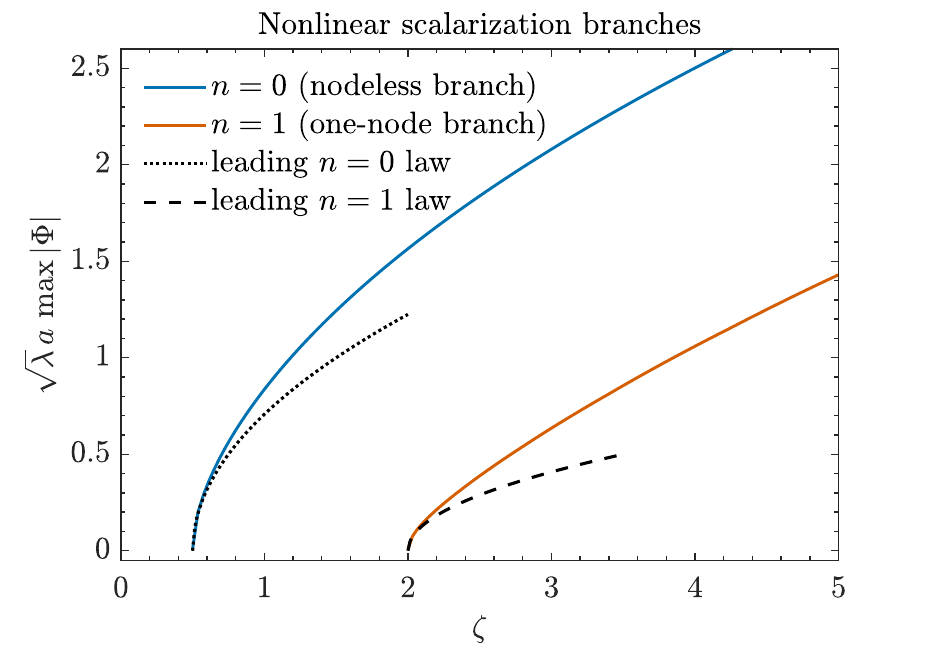}
\caption{Scalarized branches in the spherical probe-field model. The maximum field amplitude is plotted against $\zeta$ for the nodeless ($n=0$) and one-node ($n=1$) branches of Eq.~\eqref{eq:nonlinear}. The dotted and dashed curves show the leading local approximations, Eq.~\eqref{eq:pitchfork}. The figure says nothing about the stability of the supporting phantom sector or of a fully backreacted solution.}
\label{fig:branches}
\end{figure}

Figure~\ref{fig:branches} shows the two lowest branches, obtained by Newton continuation of Eq.~\eqref{eq:nonlinear}, together with the pitchfork asymptotics. Part (i) is a no-hair statement for the spherical probe-field model: below the first threshold there is no nontrivial static spherical solution. Above it, the scalar-free configuration is linearly unstable and, near onset, the nodeless branch has lower probe energy. The evolution in Fig.~\ref{fig:evol} relaxes toward this branch for the chosen initial data, although this does not prove global attraction or nonlinear stability. For $\zeta=0.75$ and $\lambda a^2=1$ the evolution settles at $\sqrt\lambda\,a\,\Phi(t,0)\to0.5629$, against $0.5627$ for the static solution. Because the field is massless, the excess energy can be radiated away to infinity.

Far from the throat the quartic term is marginal, as it always is in $3+1$ dimensions. The static equation $-u''+V_0u+\lambda u^3/r^2=0$ implies that $u=r\phi$ obeys $\ddot u-\dot u=\lambda u^3$ in the variable $\ln|x|$. As a result, the scalarized configurations decay asymptotically as $\Phi\simeq\pm[2\lambda\ln(|x|/a)]^{-1/2}|x|^{-1}$ instead of carrying a free monopole charge. The near-onset form $\Phi\simeq A_0a/|x|$ gives way to this tail at $\ln(|x|/a)\sim(\lambda a^2A_0^2)^{-1}$.

\section{Relation to the fully coupled theory}\label{sec:backreaction}

Now couple $\Phi$, with action~\eqref{eq:action}, to general relativity and to the phantom field $\chi$ that supports the Ellis geometry. The Ellis metric with its phantom profile and $\Phi\equiv0$ remains an exact solution, since every $\Phi$-dependent term in the Einstein and $\chi$ equations is at least quadratic in $\Phi$. This includes the nonminimal contribution $\zeta(G_{\mu\nu}+g_{\mu\nu}\Box-\nabla_\mu\nabla_\nu)\Phi^2$. When we linearize about this solution, $\delta\Phi$ obeys exactly Eq.~\eqref{eq:kg} on the Ellis background, decoupled from $(\delta g_{\mu\nu},\delta\chi)$. Three conclusions follow.

\emph{(a) The probe results are exact.} Sections~\ref{sec:threshold}--\ref{sec:onset} describe the exact $\Phi$ sector of the linear perturbation spectrum of the coupled theory. The thresholds, the mode counts, and the onset laws are not test-field approximations.

\emph{(b) Interpretation of the thresholds.} The values $\zeta_{n\ell}$ are exact zero-mode thresholds of the scalar block and hence candidate bifurcation points of the fully coupled system. Showing that branches of backreacted Ellis-type wormholes actually emanate from them requires solving the coupled nonlinear boundary-value problem and checking the relevant nondegeneracy conditions. The branches of Sec.~\ref{sec:nonlinear} belong to the probe-field model and should not be identified with fully backreacted solutions.

\emph{(c) What remains open.} The $(\delta g,\delta\chi)$ sector contains the known instability of the phantom-supported Ellis wormhole \cite{ShinkaiHayward2002,GonzalezLinear2009,GonzalezNonlinear2009,GonzalezCharged2009,Cremona2019,AzadEtAlPLB2024}, which $\zeta$ does not affect at linear order. Whether the backreaction of a scalarized $\Phi$ can change that, as happens for other wormhole families supported by additional fields \cite{AntoniouWH2020,BlazquezChewKunz2018,CanateSultanaKazanas2019,BronnikovReview2018}, can only be decided by constructing the backreacted branches, which we leave for future work. In the analog systems mentioned in Sec.~\ref{sec:intro} \cite{Dandoloff2010,deSouza2022,Azevedo2021,BarceloLRR2011,LeonhardtPhilbin2006} the geometry is fixed, the $(\delta g,\delta\chi)$ sector is absent, and the probe-limit analysis is complete.

\section{Conclusions}\label{sec:conclusion}

On the Ellis wormhole, curvature-induced scalarization through the Ricci coupling can be treated exactly, and since $\mathcal G\equiv0$ for this geometry, the Ricci coupling is the only quadratic-curvature channel available. Everything rests on the compact coordinate $x=a\tan z$. For static fields it produces a trigonometric P\"oschl--Teller problem with thresholds $\zeta_{n\ell}=(n+\ell+1)^2/2$. For growing modes it produces a compact Sturm--Liouville problem in which the coupling is the eigenvalue, and a monotonicity argument then gives the mode count $N_\ell(\zeta)$. For the nonlinear static problem it leads to a variational problem on a finite interval, with the same $\sec^4z$ weight.

We found two distinct ways in which unstable modes are born. At the $s$-wave thresholds, which are zero-energy resonances, the growth rate rises linearly with coefficient $\pi/[2(n+1)^2]$; at the first threshold we also obtained the second-order term, Eq.~\eqref{eq:secondorder}. At the thresholds with $\ell\geq1$, which are normalizable zero modes, the growth rate rises as a square root, and its coefficient is exactly the curvature-weighted matrix element on the threshold profile, given by finite sums of beta functions. With a positive quartic self-interaction, the spherical probe-field equation has no nontrivial finite-energy static solution for $\zeta\leq1/2$. Projection onto the threshold profiles gives the leading branch amplitudes and energy, and the time evolution we performed saturates on the nodeless static profile.

Because $\Phi$ decouples at linear order, the thresholds are exact zero-mode thresholds of the scalar block of the coupled Einstein--phantom--scalar system. They identify candidate bifurcation points but do not, by themselves, prove that fully backreacted scalarized wormholes exist or are stable. The limitations of the present analysis should be kept in mind. The Ellis background has its own instability in the phantom sector, which the scalar calculation does not remove. The quartic model is treated in the probe approximation, and its local branch coefficients follow from projection; a rigorous bifurcation theory for the singular endpoint problem in a suitable weighted function space remains to be developed. Finally, the time evolutions support saturation for the initial data tested but are not a proof of global attraction or nonlinear stability. Several extensions suggest themselves: constructing backreacted branches and studying their stability; treating the asymmetric Ellis--Bronnikov family with mass, which is no longer ultrastatic; and considering massive fields, whose spectral thresholds are modified by the mass gap. In analog realizations of the catenoidal geometry, where the geometry is externally prescribed, the probe-field results apply directly.

\emph{Data availability.} The numerical codes used to produce the figures and Table~\ref{tab:onset} are available as ancillary files with the arXiv version of this article.

\begin{acknowledgments}
C.F. acknowledges financial support from the Conselho Nacional de Desenvolvimento Cient\'ifico e Tecnol\'ogico (CNPq, Brazil) through a research productivity grant (Grant No.~311781/2021-7).
\end{acknowledgments}

\appendix

\section{Beta-function sums for the onset coefficients}\label{app:beta}

Write $p=\ell+1$ and $C_n^{p}(y)=\sum_{j=0}^{\lfloor n/2\rfloor}c_j\,y^{n-2j}$, with $c_j=(-1)^j2^{n-2j}(p)_{n-j}/[j!(n-2j)!]$. The substitution $y=\sin z$ and the identity $\int_{-1}^1y^{2q'}(1-y^2)^qdy=B(q'+\tfrac12,q+1)$ give
\begin{equation}
\mathcal I_{n\ell}(q)=\sum_{j,k=0}^{\lfloor n/2\rfloor}c_jc_k\,B\!\left(n-j-k+\tfrac12,\,q+1\right),
\end{equation}
with $I^{(+)}_{n\ell}=\mathcal I_{n\ell}(\ell+\tfrac12)$ and $I^{(-)}_{n\ell}=\mathcal I_{n\ell}(\ell-\tfrac32)$; both converge for $\ell\geq1$. Thus $s_{n\ell}=2\mathcal I_{n\ell}(\ell+\tfrac12)/\mathcal I_{n\ell}(\ell-\tfrac32)$ is a finite closed expression for all $n\geq0$ and $\ell\geq1$. For $n=0$, the integral $\int_{-\pi/2}^{\pi/2}\cos^mz\,dz=\sqrt\pi\,\Gamma(\tfrac{m+1}2)/\Gamma(\tfrac m2+1)$ leads to Eq.~\eqref{eq:s0l}. For $\ell=1,\ldots,6$ the next two families are $s_{1\ell}=\tfrac14,\tfrac58,\tfrac78,\tfrac{21}{20},\tfrac{33}{28},\tfrac{143}{112}$ and $s_{2\ell}=\tfrac{15}{136},\tfrac{7}{20},\tfrac{105}{188},\tfrac{99}{136},\tfrac{429}{496},\tfrac{715}{732}$.

\section{Numerical methods and convergence}\label{app:numerics}

\emph{Compact eigenvalue problem.} We discretize Eq.~\eqref{eq:compactkappa} by Chebyshev collocation with $N$ interior nodes and homogeneous Dirichlet conditions at $z=\pm\pi/2$. These conditions are consistent with the physical eigenfunctions, which for $\sigma>0$ decay exponentially in $|x|$ and are therefore strongly suppressed near the endpoints of the compact interval. Eigenvalues change by less than $10^{-10}$ between $N=120$ and $N=300$. Very close to an $s$-wave threshold the growing mode extends out to $|x|\sim\sigma^{-1}$, which in $z$ means within a distance of order $\sigma a$ of the endpoints; there we use up to $N=700$. At fixed $\zeta$ the growth rate is found by solving $\Lambda_{n\ell}(\sigma)=2\zeta$ with Brent's method.

\emph{Onset coefficients.} For $\ell=0$ we evaluate $\sigma a/(\zeta-\zeta_c)$ at $\zeta-\zeta_c=10^{-3}$ and $2\times10^{-3}$ and extrapolate linearly; the result differs from $\pi/[2(n+1)^2]$ by less than $2\times10^{-4}$ in relative terms. For $\ell\geq1$ we evaluate $\sigma^2a^2/(\zeta-\zeta_c)$ for $\zeta-\zeta_c$ between $10^{-2}$ and $10^{-6}$. It approaches $s_{n\ell}$ as $(\zeta-\zeta_c)^{1/2}$ for $\ell=1$ and as $\zeta-\zeta_c$ for $\ell\geq2$.

\emph{Nonlinear problem.} We solve Eq.~\eqref{eq:nonlinear} by Newton iteration on the same collocation grid ($N=160$ and $240$), continuing in $\zeta$ from the pitchfork initial guess. The throat amplitudes at the two resolutions agree to $10^{-7}$. At $\zeta-\tfrac12=5\times10^{-4}$ the energy agrees with the asymptotic value $E_0$ to $10^{-3}$ in relative terms.

\emph{Time evolution.} We integrate the $s$-wave equation with second-order finite differences in $x$ and fourth-order Runge--Kutta in $t$, with $\Delta t=h/2$. The growth rate is fitted over $40\leq t/a\leq80$.

\end{document}